\documentclass[11pt,a4paper]{article}
\usepackage[margin=1in]{geometry}
\usepackage[utf8]{inputenc}
\usepackage[T1]{fontenc}
\usepackage{amsmath,amssymb,amsthm}
\usepackage{enumitem}
\usepackage{hyperref}
\usepackage{longtable}
\usepackage{booktabs}
\usepackage{xcolor}
\usepackage{graphicx}

\hypersetup{colorlinks=true,linkcolor=blue,urlcolor=blue,citecolor=blue}

\title{A Multi-Layer AI Framework\\
for Information Landscape Analysis}

\author{
Maryam Fooladi\textsuperscript{1},
Federico Bottino\textsuperscript{1}\\[4pt]
\textsuperscript{1}Kakashi Ventures Accelerator (KVA) / Newjee
}
\date{}

\begin{document}
\maketitle

\begin{abstract}
This paper introduces a web-based AI platform for structured analysis of information disorder.
Rather than delivering simplified verdicts, the system decomposes content into eleven independently
scorable dimensions, including metadata verification, factual claim extraction, source credibility
profiling, framing analysis, and propagation detection. At the heart of its architecture sits what
we call \emph{claim--source independence}: factual veracity and source credibility run on separate
tracks, preventing the guilt-by-association reasoning that compromises most existing tools. We
demonstrate the platform through a case study of the February 2026 Russian disinformation campaign
targeting Emmanuel Macron via fabricated Epstein connections. The multi-scale architecture
(keyword-level ecosystem mapping paired with URL-level source evaluation) correctly differentiates
three qualitatively distinct information problems nested within a single topic space: active
fabrication, quality journalism about fabrication, and passive quality degradation through
unvetted republication. Each receives its own trust grade, manipulation profile, and source
assessment. We are candid about limitations, including the fundamental difficulty of building on
LLM-generated outputs without validated evaluation metrics.
\end{abstract}

\noindent\textbf{Keywords:} information disorder, disinformation detection, claim--source
independence, multi-layer analysis, LLM-based analysis, FIMI

%% ========================================================================
\section{Introduction}
%% ========================================================================

False stories on Twitter spread roughly six times faster than accurate
ones~\cite{vosoughi2018spread}. The downstream consequences are not theoretical. Allcott and
Gentzkow~\cite{allcott2017socialmedia} traced measurable distortions in electoral behavior, while
van der Linden and colleagues~\cite{vanderlinden2020inoculating} documented how disinformation
warped public health responses during COVID-19. But ``fake news'' as a framing device has always
been too blunt. Wardle and Derakhshan~\cite{wardle2017informationdisorder} proposed the
``information disorder'' framework precisely because the actual landscape is far messier than any
binary between true and false can accommodate. Their taxonomy forces a reckoning with intent,
context, and propagation pathways, all of which matter as much as whether a claim happens to be
factually correct.

Traditional fact-checking hits a wall here. Manual verification cannot match the velocity of online
content flows. Even if it could, binary verdicts leave enormous analytical territory unexplored.
Consider the statement that is technically accurate but strategically omits the one detail that
would reverse its meaning. Or the legitimate photograph placed in a fabricated context. These are
not edge cases. They are, increasingly, the norm.

This paper presents a web-based AI platform that approaches the problem differently. Instead of
delivering verdicts, the system decomposes content into independently scorable dimensions across
eleven analytical layers. The central architectural decision is \emph{claim--source independence}:
factual veracity and source credibility are scored on separate tracks, preventing
guilt-by-association reasoning where disreputable outlets' accurate reporting gets dismissed and
prestigious broadsheets' misleading framing gets a pass. The platform operates in two modes:
URL-based deep analysis of individual articles and keyword-based narrative mapping across sources.
Both produce transparent trust scores aimed at journalists, researchers, and policy analysts who
need to understand information ecosystems rather than receive simplified judgments about them.

One clarification is important. The platform is not a primary disinformation detector. It does not
crawl bot networks or perform technical attribution. It operates downstream, synthesizing publicly
available reporting about information operations into structured, multi-layered analytical
outputs. This distinction shapes how every subsequent result should be interpreted.

%% ========================================================================
\section{Related Work}
%% ========================================================================

Research on computational approaches to disinformation has expanded rapidly, spanning automated
fact-checking, credibility assessment, and narrative analysis. We focus on threads most directly
relevant to our contribution; comprehensive surveys are available in Zeng et
al.~\cite{zeng2021automated} and Guo et al.~\cite{guo2022survey}.

Automated fact-checking systems, from ClaimBuster~\cite{hassan2017claimbuster} through
FEVER~\cite{thorne2018fever}, treat verification as classification: claims mapped to labels of
true, false, or intermediate. This works for straightforward assertions but lacks vocabulary for
subtler pathologies of information disorder. What do you do with the accurate statement that
strategically withholds reversing context? Framing-as-classification has a ceiling, and it is
lower than we might like.

Source credibility assessment has taken two paths: human curation (NewsGuard, Media Bias/Fact
Check) and automation~\cite{baly2018predicting, gruppi2021nelagt}. The persistent limitation is
that source credibility and claim veracity keep getting entangled. A tabloid can publish careful
investigative work. A prestigious broadsheet can frame facts misleadingly. Remarkably few systems
treat disentangling these dimensions as a first-order design requirement.

Applied platforms including Bellingcat, VeraAI~\cite{lupi2023veraai}, and
AskVera~\cite{wuhrl2023askvera} have operationalized research for journalist workflows. These
represent real progress. Yet most default to verdict delivery over analytical transparency:
telling users what to think about content rather than showing them what is happening in its
information ecosystem.

Our platform addresses these gaps simultaneously. Unlike claim-verification systems, it analyzes
multiple dimensions without forcing content through a classification bottleneck. Unlike
source-rating tools, it explicitly separates claim veracity from source credibility as a concrete
scoring mechanism. Unlike tools built exclusively for professional fact-checkers, it makes
structured analysis accessible to broader audiences, including researchers investigating
cross-cultural information dynamics~\cite{dabbous2022fakenews} and the challenges of annotator
disagreement in disinformation labeling~\cite{plank2022humanlabel}.

%% ========================================================================
\section{System Architecture}
%% ========================================================================

\subsection{Overview and Transparency}

The platform runs on Next.js~14 with a Supabase PostgreSQL backend. The analytical core uses
Claude Sonnet~4 (Anthropic), augmented with real-time web search; 20 queries for main analysis,
8 for propagation tracking. Topic classification runs on GPT-4o-mini (OpenAI). These three
pipelines execute concurrently.

We should be transparent: this is an LLM-based platform, not a traditional NLP pipeline with
trained classifiers and evaluation metrics on held-out test sets. The analytical outputs come from
structured prompting. This has implications for interpretability and validation that we address
in Section~\ref{sec:discussion}.

\subsection{Analysis Modes}

\paragraph{URL-based analysis} performs deep examination of individual articles, extracting
claims, evaluating source health, analysing framing and metadata.

\paragraph{Keyword-based analysis} aggregates across multiple sources to map the narrative
landscape around a topic.

Their relationship is deliberate: keyword analysis provides the ecosystem-level picture, URL
analysis provides source-level granularity. Running both on the same topic produces different
outputs, and they are meant to be different.

\subsection{Analytical Framework}

The framework decomposes content into eleven independently scored layers:

\begin{enumerate}[leftmargin=*,itemsep=2pt]
  \item \textbf{Metadata Assessment}, examining publication date, authorship, technical
  authenticity signals.
  \item \textbf{Factual Claims Analysis}, with multi-source triangulation.
  \item \textbf{Source Health Scan}, evaluating publisher credibility along six weighted axes
  (Reputation 20\%, Factual Consistency 30\%, Transparency 15\%, Bias History 5\%, Manipulation
  History 20\%, Fact-Check Collaboration 10\%).
  \item \textbf{Narrative Analysis}, tracking information organization and omissions.
  \item \textbf{Media Integrity Assessment}, targeting image and video manipulation.
  \item \textbf{Language and Ideological Analysis}, drawing on computational pragmatics to surface
  how language choices shape interpretation independently of factual content.
  \item \textbf{Framing Detection}, maintaining strict separation between descriptive
  \emph{Narrative Frames} (characterizing which strategies are employed) and quantitative
  \emph{Manipulation Indicators} (assessing how strongly manipulation patterns manifest).
  \item \textbf{Propagation Analysis}, tracking spread patterns with outlet-frame mapping.
  \item \textbf{User Interaction Prompts}.
  \item \textbf{Reliability Summary}.
  \item \textbf{Suggested Actions}.
\end{enumerate}

The weighting rationale follows established practice. We weight Factual Consistency highest at
30\% because historical accuracy is the strongest predictor of future reliability. Bias History
receives only 5\%: editorial perspective does not inherently compromise factual accuracy. A source
can be openly partisan and scrupulously accurate at the same time.

\subsection{Claim--Source Independence}

This is the platform's most consequential design decision. Conventional fact-checking conflates
source reputation with content accuracy, and both directions produce errors. The system flags
\emph{high-truth, low-source} (HTLS) scenarios, triggered when claim veracity $\geq 0.80$ and
source reliability $\leq 0.40$. Rather than averaging these into a comfortable middle score, the
system surfaces the tension explicitly. The inverse case receives analogous treatment.

\subsection{Scoring and Output}

Trust scores map onto four tiers: A (75--100\%), B (50--74\%), C (25--49\%), D (0--24\%).
Calibration caps prevent overconfidence: even strong outlets cannot exceed approximately 0.88.
Results appear across tabbed views (Facts, Source, Citations, Verdict, Framing, Propagation,
Actions, Entities), each with transparent explanations.

%% ========================================================================
\section{Case Study: The Macron--Epstein Disinformation Campaign}
%% ========================================================================

\subsection{Case Selection}

In early February 2026, Russian-linked networks launched a Foreign Information Manipulation and
Interference (FIMI) operation falsely linking President Macron to convicted sex offender Jeffrey
Epstein. The campaign piggybacked on the legitimate release of Epstein court documents by the
U.S.\ Department of Justice (January 30, 2026), mixing real document references with fabricated
emails and AI-generated video. France's Viginum agency attributed the operation to networks
Storm-1516 and Matryoshka; Bot Blocker and outlets including Euronews, France24, and Reuters
independently confirmed attribution.

We selected this case because it represents documented, attributed FIMI with ground truth, deploys
multiple disorder types simultaneously, and has European focus and recency aligning with this
community's priorities. Critically, the analysis used the keyword query ``Macron Epstein'' without
terms like ``disinformation'' or ``Russia,'' testing whether the system could surface information
disorder unprompted. The analysis ran on February 13, 2026.

\subsection{Keyword Ecosystem Analysis}

The system assigned Trust Grade~D (18\%), with Claim Veracity at 5\% and Source Reliability at
85\%. The divergence matters. The reporting sources (Euronews, France24, Reuters) are highly
credible, but the claims circulating in the ecosystem (fabricated emails, doctored documents,
AI-generated video) are almost entirely false. The trust formula correctly produces a low
composite grade reflecting ecosystem-level falsity rather than inflating the score based on
debunking source quality.

The Facts layer extracted seven verified claims, including Viginum's detection and attribution,
distribution through a cloned France-Soir website (\texttt{france-soir.net} impersonating
\texttt{france-soir.fr}), and identity theft of \emph{Le Parisien} journalist Victor Cousin.

The Framing layer identified three Narrative Frames (\emph{Scapegoating}, \emph{False
Victimization}, \emph{Moral Panic}) alongside five Manipulation Indicators at high confidence:
\texttt{fabricated\_news} (95\%), \texttt{disinformation\_campaign} (90\%),
\texttt{imposter\_source} (85\%), \texttt{manipulated\_content} (80\%),
\texttt{influence\_operation} (90\%).

The Propagation layer proved particularly revealing. The Outlet-Frame Mapping
(Figure~\ref{fig:outlet-frame}) identified three clusters: (1) an \emph{Official
Counter-Disinformation Frame} (Euronews, France24, Reuters), (2) a \emph{Technical Analysis Frame}
(UNITED24 Media, The Insider, Bot Blocker), and (3) a \emph{Sensationalist Amplification Frame}
(Pravda EN, fake France-Soir clones). The ecosystem is tripartite, not binary. Any classifier
forcing a two-way split would miss this.

\begin{figure}[h]
\centering
\fbox{\parbox{0.85\textwidth}{\centering\vspace{1em}
\textit{[Figure placeholder: Outlet-Frame Mapping panel listing the three frame clusters
(Official Counter-Disinformation, Technical Analysis, Sensationalist Amplification) with their
associated outlets and characteristic language, plus the propagation pathway timeline.]}
\vspace{1em}}}
\caption{Outlet-Frame Mapping showing three frame clusters with associated outlets and
characteristic language.}
\label{fig:outlet-frame}
\end{figure}

The \textbf{Entity Resolution} of Victor Cousin went beyond standard NER, identifying Cousin as a
\emph{victim} of identity theft by cross-referencing factual claims, source analysis, and
propagation patterns simultaneously.

\subsection{Multi-Scale Comparison}

To test multi-scale differentiation, we ran URL-level analyses of two sources within the same
ecosystem.

\paragraph{Euronews} received Trust Grade~A (86\%), Claim Veracity 92\%, Source Reliability 85\%.
This is fundamentally different from the keyword analysis's D (18\%), and it should be. The
keyword analysis evaluates the \emph{ecosystem of claims} (dominated by fabrication); the URL
analysis evaluates \emph{one article's reporting} (well-sourced journalism). Same topic, different
scale, different valid outputs. The Manipulation Indicators confirmed the distinction:
\texttt{reliable\_reporting} (85\%), \texttt{government\_source\_reliance} (75\%),
\texttt{anti\_disinformation\_narrative} (70\%).

\paragraph{IndoPremier} covered the legitimate Fabrice Aidan diplomat case (a real French
diplomat named in authentic Epstein files). Trust Grade: B (72\%), Claim Veracity 82\%, Source
Reliability 35\%. The system triggered an \textbf{HTLS conflict flag}: ``reliable information from
a questionable source.'' IndoPremier's claims scored high because they originate from AFP wire
reporting; the source scored low because it lacks editorial infrastructure for independent
verification. A qualitatively different problem from both ecosystem-level fabrication and quality
original reporting. Manipulation Indicators confirmed this third profile:
\texttt{republished\_content} (85\%), \texttt{limited\_verification} (70\%),
\texttt{neutral\_framing} (60\%).

\subsection{Synthesis}

Figure~\ref{fig:three-scale} summarizes the three-scale comparison.

\begin{figure}[h]
\centering
\fbox{\parbox{0.85\textwidth}{\centering\vspace{1em}
\textit{[Figure placeholder: grouped bar chart of Source Reliability vs.\ Claim Veracity across
three scales --- Keyword Ecosystem (Trust~D, 18\%; SR~85\%, CV~5\%), Euronews~URL (Trust~A, 86\%;
SR~85\%, CV~92\%), IndoPremier~URL (Trust~B, 72\%; SR~35\%, CV~82\%) --- annotated with the trust
grade above each group.]}
\vspace{1em}}}
\caption{Comparative analysis of Source Reliability and Claim Veracity across three analytical
scales within the ``Macron Epstein'' information ecosystem. Trust grades shown above each group.
The Keyword Ecosystem analysis reveals a sharp divergence between high Source Reliability (85\%)
and near-zero Claim Veracity (5\%), reflecting ecosystem-level falsity despite credible reporting
sources. The IndoPremier analysis triggers a high-truth, low-source (HTLS) conflict flag, with
Claim Veracity (82\%) substantially exceeding Source Reliability (35\%).}
\label{fig:three-scale}
\end{figure}

The same topic contains three qualitatively distinct information problems, and the architecture
correctly differentiates all three. Keyword analysis alone would not reveal that sources range
from excellent to questionable. URL analyses alone would miss the ecosystem-level FIMI operation.
The scales are complementary, not redundant.

The Trust Score paradox---D grade despite 85\% Source Reliability---illustrates what happens when
source quality and claim quality are honestly integrated. The Framing dual structure allows
describing \emph{how} actors frame issues independently from assessing \emph{whether} manipulation
is present. The Outlet-Frame Mapping reveals a tripartite ecosystem invisible to binary bias
classifiers. The Entity Resolution of Victor Cousin demonstrates what becomes possible when
multiple layers converge on the same entity.

%% ========================================================================
\section{Discussion and Limitations}
\label{sec:discussion}
%% ========================================================================

\paragraph{Analytical positioning.}
The platform did not discover the Russian bot networks. Viginum and Bot Blocker did that work.
What the platform did was synthesize fragmented public reporting into structured outputs (frame
clusters, propagation pathways, severity-graded patterns, cross-referenced entity profiles) that
would otherwise require considerable manual effort. The value proposition is structuring
intelligence, not generating it.

The multi-scale finding is the paper's central contribution. What appears to be a single topic
actually contains three qualitatively distinct information problems: active fabrication,
responsible reporting about fabrication, and passive quality degradation. This validates offering
both ecosystem-level and source-level analysis. The case study is itself inherently cross-cultural:
a Russian operation targeting a French president, reported by European and Turkish outlets, with an
Indonesian website republishing French wire content. The platform's English-language retrieval
captured only one slice of this multilingual landscape, underscoring the urgency of multilingual
expansion for any system analyzing European FIMI operations where campaigns exploit language
boundaries~\cite{dabbous2022fakenews}.

\paragraph{Limitations.}
Every quantitative output (claim veracity, source reliability, manipulation indicators, trust
grades) comes from Claude Sonnet~4 through structured prompting. None have been validated against
ground truth annotations or benchmarked against expert judgments at scale. The scores produce
plausible orderings, but calibration properties are unknown. A 92\% veracity score does not carry
the same meaning as a 92\% F1 from a validated classifier. The source health weights are
heuristic. The system retrieves only English-language sources. Building on an LLM means inheriting
its failure modes.

A preliminary evaluation by two computational linguistics researchers showed 67\% agreement across
nine assessment dimensions, with full agreement on the keyword ecosystem and disagreements on
cases where claim--source independence creates analytical tensions. These disagreements illustrate
the human-label variation problem~\cite{plank2022humanlabel} applied to information disorder
assessment: reasonable experts applying different weighting intuitions to identical outputs.

%% ========================================================================
\section{Conclusion}
%% ========================================================================

We have presented a multi-layer AI platform for structured analysis of information disorder at
two complementary scales. The case study demonstrated that a single topic space can harbor
qualitatively distinct information problems, each correctly characterized with distinct trust
grades, manipulation profiles, and source assessments. The platform occupies a specific niche:
structuring fragmented reporting into multi-layered outputs that support human analysis. Its
limitations---unvalidated confidence scores, English-language bias, heuristic weights, no user
studies---define the research agenda: multilingual retrieval, longitudinal tracking, empirical
validation, and structured user studies with information professionals.

%% ========================================================================
\section*{Ethical Considerations}
%% ========================================================================

This work raises several ethical questions that we want to address directly.

The platform relies on commercial LLMs (Claude Sonnet~4 and GPT-4o-mini) for its analytical core.
This creates a dependency on opaque, proprietary systems whose internal reasoning cannot be fully
audited. Users of the platform must understand that trust scores, manipulation indicators, and
veracity assessments are probabilistic outputs generated through structured prompting, not
deterministic classifications from validated models. We have attempted to mitigate potential
overreliance by making the scoring transparent and by presenting analytical decompositions rather
than binary verdicts.

The case study involves a real, documented disinformation campaign. All information referenced in
this paper comes from publicly available reporting by established outlets (Euronews, France24,
Reuters) and official government sources (Viginum). We do not reproduce or amplify any of the
fabricated content from the campaign itself. The identity theft of journalist Victor Cousin is
discussed because it was already publicly reported and attributed; we include it to demonstrate
the platform's entity resolution capabilities, not to further expose the victim.

There is a broader risk that platforms performing automated information analysis could themselves
become tools of censorship or political manipulation if deployed without safeguards. Our system
is designed to support human analysts, not to replace editorial judgment or automate content
moderation decisions. We strongly discourage any use of this platform for automated content
removal or suppression without human oversight.

Finally, the platform currently operates only in English, which introduces a linguistic bias in
its analytical outputs. Information ecosystems are inherently multilingual, and evaluating them
through a single language inevitably produces an incomplete picture. We acknowledge this
limitation and consider multilingual expansion a priority for responsible development.

%% ========================================================================

\end{document}